\documentclass[aps]{revtex4-2}
\usepackage{booktabs}
\usepackage[utf8]{inputenc}
\usepackage{amssymb}
\usepackage{amsfonts}
\usepackage{xcolor}
\usepackage{amsmath,amssymb,graphicx}
\usepackage{graphicx}% Include figure files
\usepackage[pdftex]{hyperref}
\usepackage{dcolumn}% Align table columns on decimal point
\usepackage{bm}% bold math
\usepackage{tikz}

\def\be{\begin{equation}}
\def\ee{\end{equation}}
\def\bea{\begin{eqnarray}}
\def\eea{\end{eqnarray}}

\usepackage{pstricks}
\begin{document}
\title{CP Violation in $D \to KK$ Decays: A Comparative Analysis of Triplet and Sextet Diquarks}

%\title{Enhanced CP Asymmetry in $D^0 \to K_S^0 K_S^0$ from Scalar Diquarks }

\author{David Delepine}
\email{delepine@fisica.ugto.mx}
\affiliation{{\fontsize{10}{10}\selectfont{Division de Ciencias e
Ingenierias,  Universidad de Guanajuato, C.P. 37150, Le\'on,
Guanajuato, M\'exico.}}}

%\author{ Gaber
%Faisel}\email{gaberfaisel@sdu.edu.tr}
%\affiliation{{\fontsize{10}{10}\selectfont{Department of Physics,
%Faculty of Engineering and Natural Sciences, S\"uleyman Demirel
%University, Isparta, Turkey 32260 .}}}

\author{Shaaban Khalil}
\email{skhalil@zewailcity.edu.eg}
\affiliation{{\fontsize{10}{10}\selectfont{Centre for Theoretical Physics, Zewail City of Science and Technology, 6th October City, 12588, Giza, Egypt.}}}

\author{ Carlos A. Ramirez}
\email{jpjdramirez@yahoo.com}
\affiliation{{\fontsize{10}{10}\selectfont{Depto. de Fisica,
Universidad de los Andes, A. A. 4976-12340, Bogot\'a, Colombia.}}}

\date{\today}

%%%%%%%%%%%%%%%%
\begin{abstract}
%%%%%%%%%%%%%%%%

Recent measurements of the CP asymmetry in the decay $D^0 \rightarrow K_S^0 K_S^0$ by the CMS collaboration, $A_{CP}(K_S^0 K_S^0) = (6.2 \pm 3.0 \pm 0.2 \pm 0.8)\%$, and by LHCb, $A_{CP}(D^0 \to K_S^0 K_S^0) = (1.86 \pm 1.04 \pm 0.41)\%$, suggest possible deviations from Standard Model (SM) expectations, which predict asymmetries below the percent level. This singly Cabibbo-suppressed decay is particularly sensitive to new physics, as the leading amplitudes vanish in the exact U-spin symmetry limit and the process is dominated by W-exchange topologies. We investigate scalar diquark contributions to this decay, comparing color-sextet and color-triplet representations. We find that the color-sextet diquark, characterized by a symmetric color structure $(C_1^{\mathrm{NP}} = C_2^{\mathrm{NP}})$, avoids color suppression and can generate CP asymmetries in the range $0.5\%$--$1.5\%$ for a diquark mass of order 1~TeV. In contrast, the color-triplet contribution is strongly suppressed due to destructive interference from its antisymmetric color structure. We further show that a flavor hierarchy in the sextet couplings, with $\lambda_{ud} > \lambda_{us}$, can simultaneously account for the observed deviation from the U-spin sum rule in $D^0 \to K^+ K^-$ and $D^0 \to \pi^+ \pi^-$ and the measured CP asymmetry in $D^0 \to K_S^0 K_S^0$. These results identify color-sextet scalar diquarks as viable candidates for explaining enhanced CP violation in charm decays.

\end{abstract}
\maketitle
%%%%%%%%%%%%%%%%%%%%%%%%%%%%%%%%%%%%%%
\section{Introduction}

Charge–parity (CP) violation is one of the most profound phenomena in particle physics, playing a central role in our understanding of the matter–antimatter asymmetry of the Universe. Within the Standard Model (SM), CP violation originates from a single irreducible complex phase in the Cabibbo–Kobayashi–Maskawa (CKM) quark mixing matrix. While this mechanism successfully explains the observed CP violation in the kaon and beauty sectors, its effects in the charm sector are predicted to be highly suppressed. This suppression arises from the hierarchical structure of the CKM matrix, the Glashow–Iliopoulos–Maiani (GIM) mechanism, and the dominance of light quark contributions in charm transitions. As a consequence, direct CP asymmetries in singly Cabibbo-suppressed (SCS) charm decays are generally expected to be at most of order $10^{-3}$ within the SM.

The experimental study of CP violation in charm decays has recently entered a new era of precision, driven by the large datasets collected by the LHCb and CMS experiments. Of particular interest is the decay channel $D^0 \to K_S^0 K_S^0$, which provides a uniquely sensitive probe of new sources of CP violation. Recent measurements by the LHCb collaboration have reported a direct CP asymmetry \cite{LHCb:2025ezf, LHCb:2019hfv}: 
\[
A_{CP}(D^0 \to K_S^0 K_S^0) = (1.86 \pm 1.04 \pm 0.41)\%,
\]
while CMS has reported a consistent result at the percent level. These values are significantly larger than the SM expectation, which is usually estimated to lie below $0.5\%$ due to the strong suppression of CP-violating amplitudes in this channel\cite{CMS:2024hsv}.

The decay $D^0 \to K_S^0 K_S^0$ is a very sensitive probe for New Physics (NP)\cite{Nierste:2015ksks}. In the SM, the leading tree-level amplitude vanishes in the exact U-spin limit, leaving the decay to proceed through suppressed U-spin-breaking effects and loop-induced penguin diagrams\cite{Grossman:2006jg,Cheng:2012wr}. This inherent suppression minimizes the ``background'' of SM CP violation, making the channel an ideal laboratory for identifying new sources of CP violation\cite{Nierste:2015ksks}.

Theoretical interest in charm CP violation has been further stimulated by recent experimental indications of deviations from U-spin sum rules relating the CP asymmetries of the SCS decay modes $D^0 \to K^+K^-$ and $D^0 \to \pi^+\pi^-$. Global fits to experimental data suggest a violation of these relations at the level of approximately $2.7\sigma$, with both asymmetries observed to be positive, contrary to the leading-order SM expectation \cite{PhysRevLett.122.211803,PhysRevLett.131.091802,LHCb:2022}.  These observations  point toward the intriguing possibility of new physics contributions to charm decay amplitudes.

%In this work, we investigate scalar diquarks as a concrete NP scenario capable of generating percent-level CP asymmetries\cite{Chen:2012ww}. By comparing the color-sextet $(\bar{\mathbf{6}}, \mathbf{1}, 1/3)$ and color-triplet $(\mathbf{3}, \mathbf{1}, -1/3)$ representations, we demonstrate that the unique color symmetry of the sextet allows it to bypass standard suppression mechanisms\cite{Chen:2012ww,Fajfer:2012nr}. We show how a hierarchy in diquark couplings can simultaneously resolve the $2.7\sigma$ tension observed in the U-spin sum rule for $D^0 \to K^+ K^-$ and $D^0 \to \pi^+ \pi^-$ decays\cite{LHCb:2022} while providing a natural explanation for the magnified signals in $D^0 \to K_S^0 K_S^0$.

Among the various BSM scenarios proposed to explain enhanced CP violation in charm decays, scalar diquarks provide a particularly compelling framework . Scalar diquarks are easily introduced in many well-motivated extensions of the SM, including grand unified theories \cite{Chen:2012ww}. based on gauge groups such as $SO(10)$ and $E_6$, composite models, and theories with extended color sectors. These states transform as colored scalars under $SU(3)_C$ and can couple directly to quark pairs, generating new four-quark interactions at tree level. Unlike SM contributions, which generate primarily vector and axial-vector operators, scalar diquark exchange can produce scalar and tensor operators, introducing new possible interference terms  and CP-violating phases.

A key feature of scalar diquarks is how they transform under the color gauge group, which determines their interference properties with SM amplitudes. Specifically, scalar diquarks transforming as color sextets possess a symmetric color structure that allows constructive interference with SM exchange amplitudes, while color-triplet diquarks, characterized by antisymmetric color contractions, generating destructive interferences and are therefore less effective in enhancing CP asymmetries. This difference between sextet and triplet scalar diquarks has important phenomenological consequences for charm decays dominated by exchange topologies, such as $D^0 \to K_S^0 K_S^0$.

In this work, we perform a detailed and systematic investigation of scalar diquark contributions to CP violation in charm decays, with a primary focus on the $D^0 \to K_S^0 K_S^0$ channel. We consider both color-sextet and color-triplet scalar diquark representations and  their effects on the effective weak Hamiltonian governing charm transitions is studied. We derive the resulting Wilson coefficients, examine the induced operator structures, and estimate the resulting CP asymmetries.  The implications of flavor-dependent diquark couplings and their ability to  explain the observed pattern of CP violation across multiple charm decay channels is explored.

Our analysis demonstrates that color-sextet scalar diquarks with masses in the TeV range and perturbative couplings is able to generate CP asymmetries at the percent level, within the range of  current experimental measures. In contrast, the color-triplet representation is suppressed due to destructive color interference. We also show that a hierarchical structure of diquark couplings can account for the observed violation of U-spin sum rules and provide an explanation of CP violation in both neutral and charged kaon final states.

This paper is organized as follows. In Section II, we review the Standard Model contributions to the decay $D^0 \to K_S^0 K_S^0$ and discuss the origin of CP violation in this channel. In Section III, we introduce the scalar diquark framework and derive the effective operators generated by diquark exchange. Section IV presents a detailed phenomenological analysis of scalar diquark contributions to CP asymmetries in charm decays, including a comparison between color-sextet and color-triplet scenarios. In section V, we discussed the experimental and theoretical constraints on diquarks models.  Finally, Section VI summarizes our results and discusses their implications for future experimental searches and theoretical developments.
%%%%%%%%%%%%%%%%%%%%%%%%%%%%%%%%%%%%%%%%
\section{Standard Model contribution}
%%%%%%%%%%%%%%%%%%%%%%%%%%%%%%%%%%%%%%%%%%%
In the Standard Model (SM), the decay $D^{0}\rightarrow K_{S}^{0}K_{S}^{0}$ is a Singly Cabibbo-Suppressed (SCS) process. Unlike Cabibbo-favored decays which are dominated by tree-level spectator diagrams, this channel proceeds primarily through non-spectator topologies: the $W$-exchange ($E$) and Penguin Annihilation ($PA$) diagrams\cite{Nierste:2015ksks}.
The effective Hamiltonian for the SCS charm decay is\cite{Grossman:2006jg,Brod:2012ud}:$$\mathcal{H}_{eff} = \frac{G_F}{\sqrt{2}} \sum_{q=d,s} \lambda_q \left( C_1 Q_1^q + C_2 Q_2^q + \sum_{i=3}^6 C_i Q_i \right)$$Where $Q_1, Q_2$ are current-current (tree) operators and $Q_{3 \dots 6}$ are QCD penguin operators.$\lambda_q \equiv V_{cq}^* V_{uq}$ and $C_i$ are the Wilson coefficients. 

\begin{eqnarray}
    Q_1^q &=& (\bar{u}_{\alpha} \gamma^{\mu} P_L q_{\beta}) (\bar{q}_{\beta} \gamma_{\mu} P_L c_{\alpha})\\
    Q_2^q &=& (\bar{u}_{\alpha} \gamma^{\mu} P_L q_{\alpha}) (\bar{q}_{\beta} \gamma_{\mu} P_L c_{\beta}) \\
    Q_3 &=& (\bar{u}_{\alpha} \gamma^{\mu} P_L c_{\alpha}) \sum_{q'} (\bar{q}'_{\beta} \gamma_{\mu} P_L q'_{\beta})\\
    Q_4 &=& (\bar{u}_{\alpha} \gamma^{\mu} P_L c_{\beta}) \sum_{q'} (\bar{q}'_{\beta} \gamma_{\mu} P_L q'_{\alpha})\\
    Q_5  & =& (\bar{u}_{\alpha} \gamma^{\mu} P_L c_{\alpha}) \sum_{q'} (\bar{q}'_{\beta} \gamma_{\mu} P_R q'_{\beta}) \\ 
    Q_6 &=& (\bar{u}_{\alpha} \gamma^{\mu} P_L c_{\beta}) \sum_{q'} (\bar{q}'_{\beta} \gamma_{\mu} P_R q'_{\alpha})
\end{eqnarray}
The usual $O_{7-10}$ are generated in SM through electroweak loop penguin and are strongly suppressed. It is why we should not include them in that section. 
Using standard notation, the full amplitude $A(D^0 \to K_S^0 K_S^0)$ is expressed as\cite{Nierste:2015ksks}:

\begin{equation}
A(D^0 \to K_S^0 K_S^0) = \frac{G_F}{\sqrt{2}} \left[ \lambda_{sd} A_{sd} - \frac{\lambda_b}{2} A_b \right]
\end{equation}
with $\lambda_{sd} \equiv (\lambda_s - \lambda_d)/2$
\begin{equation}
A_{sd} = \sum_{i=1,2} C_i \left( \langle K_S K_S | Q_i^s | D^0 \rangle - \langle K_S K_S | Q_i^d | D^0 \rangle \right)
\end{equation}

The CP-Violating Piece ($A_b$) is given by:
\begin{equation}
A_b = \sum_{i=1}^{2} C_i \langle Q_i^{u} \rangle + \sum_{i=3}^{6} C_i \langle Q_i \rangle 
\end{equation}
\begin{itemize}
    \item Term 1 ($\sum_{i=1}^2 C_i \langle Q_i^u \rangle$): Represents the Tree/Exchange diagrams.  It represents the "flavor-symmetric" part of the exchange\footnote{While the physical decay involves strange ($s$) and down ($d$) quarks, theorists use the $u$-quark operator as a reference point to isolate the flavor-symmetric part of the exchange diagram\cite{Nierste:2015ksks}}.
    \item Term 2 ($\sum_{i=3}^6 C_i \langle Q_i \rangle$): Represents the QCD Penguin loops.
\end{itemize}
\begin{table}[h!]
\centering
\caption{Effective Hamiltonian Operator Groups within the Standard Model for $D^0 \to K_S^0 K_S^0$. The operator basis and their phenomenological roles follow the standard $\Delta C = 1$ framework \cite{Buchalla:1995vs, Grossman:2006jg, Brod:2012ud}}
\label{tab:d0_operators}
\begin{tabular}{|l|l|l|l|}
\hline
\textbf{Operator Group} & \textbf{Type} & \textbf{Physical Process} & \textbf{Role in $D^0 \to K_S^0 K_S^0$} \\ \hline
$Q_1, Q_2$ & Current-Current & $W$-Exchange & Dominant; determines the Branching Ratio. \\ \hline
$Q_3 \dots Q_6$ & QCD Penguins & Gluon Loops & Main source of SM CP Violation. \\ \hline
$Q_7 \dots Q_{10}$ & EW Penguins & Photon/$Z$ Loops & Isospin-violating; sensitive to heavy NP. \\ \hline
\end{tabular}
\end{table}
Neglecting higher-order terms in the ratio $|\lambda_b / \lambda_{sd}|$, the direct CP asymmetry is computed as \cite{Nierste:2015ksks}:

$$a_{CP}^{dir} = \text{Im} \left( \frac{\lambda_b}{\lambda_{sd}} \right) \cdot \text{Im} \left( \frac{A_b}{A_{sd}} \right)$$

\begin{itemize}
    \item Weak Factor: The first term, $\text{Im} \left( \frac{\lambda_b}{\lambda_{sd}} \right) \approx -6 \times 10^{-4}$, is a fixed Standard Model value determined by CKM matrix elements.
    \item Strong Factor: The second term, $\text{Im} \left( \frac{A_b}{A_{sd}} \right) = \left| \frac{A_b}{A_{sd}} \right| \sin \delta$, depends on the magnitude of the amplitude ratio and the relative strong phase ($\delta$) between them. In reference \cite{Nierste:2015ksks}, they suggested a ratio of $|A_b / A_{sd}| \approx 3 \text{--} 5$.
\end{itemize}
In SM, a numerical estimate of the direct CP violating asymmetry in this channel is given by

$$|a_{CP}^{dir}| \approx 6 \times 10^{-4} \times (3 \text{ to } 5) \times \sin\delta \approx \mathbf{0.2\% \text{ to } 0.4\%}$$
%%%%%%%%%%%%%%%%%%%%%%%%%%%%%%%%%%%%%%%%%%%%%%%%%%%%%%%%%%%%
\section{Scalar Diquarks in Beyond the Standard Model Theories}
%%%%%%%%%%%%%%%%%%%%%%%%%%%%%%%%%%%%%%%%%%%%%%%%%%%%%%%%%%%%

%\subsection{Classification and Selection of the Diquark Representation}

%The decay $D^0 \to K_S^0 K_S^0$ is a particularly sensitive probe of New Physics (NP) due to its strong suppression in the Standard Model (SM) and the absence of dominant tree-level contributions. This makes it an ideal laboratory to search for new sources of CP violation. 
In this work, we investigate the impact of scalar diquark exchange as a concrete and well-motivated NP scenario capable of generating sizeable CP-violating effects in this channel.

The classification of scalar diquarks based on their transformation properties under the Standard Model gauge group yields eight possible representations \cite{Giudice:2011ak}. However, the specific phenomenology of the $D^0 \to K_S^0 K_S^0$ decay imposes stringent theoretical and experimental constraints, which eliminate the majority of these candidates\cite{Golowich:2007ka, Cheng:2012wr}.

In order to generate a sizeable NP contribution to the CP asymmetry in this channel, the scalar diquark model must satisfy the following four simultaneous conditions:

\begin{enumerate}
    \item \textbf{Simultaneous Coupling to Up- and Down-Type Quarks:}  
    The scalar must couple to both up-type and down-type quarks in order to mediate the required $c \to u s \bar{s}$ transition through diquark exchange \cite{Chen:2012ww}.
    
    \item \textbf{Chirality Mixing:}  
    To generate scalar and tensor four-quark operators (in particular $C_7$ and $C_8$) and enable the chiral interference mechanism responsible for CP violation, the diquark must couple to both left-handed quark doublets $Q_L$ and right-handed singlets $u_R, d_R$\cite{Chen:2012ww}.
    
    \item \textbf{Flavor Structure:}  
    The couplings in the up-quark sector must be symmetric in flavor space to avoid dangerous flavor-changing neutral currents (FCNCs) and to allow the required interference pattern \cite{Giudice:2011ak}.
    
    \item \textbf{Consistency with Flavor Constraints:}  
    The model must respect current experimental bounds, in particular those from neutral meson mixing. Crucially, the diquark must not induce tree-level contributions to $B_s - \bar{B}_s$ mixing, which severely constrains many diquark and leptoquark scenarios \cite{Dorsner:2016wpm}.
\end{enumerate}

Imposing these conditions directs us to scalar fields transforming as weak isospin singlets with hypercharge $Y=1/3$, which can exist in two viable color representations: the color anti-triplet and the color sextet \cite{Giudice:2011ak}. These correspond to scalar fields transforming as:
\begin{equation}
\phi_3 \sim (\mathbf{\bar{3}}, \mathbf{1}, 1/3)_{SM}, \quad \text{and} \quad \phi_6 \sim (\mathbf{6}, \mathbf{1}, 1/3)_{SM}
\end{equation}
under $SU(3)_C \times SU(2)_L \times U(1)_Y$.Unlike other representations which couple exclusively to right-handed quarks (e.g. $d_R d_R$) or purely to left-handed doublets, these representations allow simultaneous couplings to both chiralities. The corresponding interaction Lagrangian can be generalized for both cases as:

\begin{equation}
\mathcal{L} = \left( \frac{1}{2} \tilde{\lambda}_{ij}^{L} (\overline{Q}_{i}^{I,\alpha})^{c} Q_{j}^{J,\beta} \epsilon_{IJ} + \tilde{\lambda}_{ij}^{R} \overline{u}_{i}^{\alpha c} d_{j}^{\beta} \right) \Gamma_{\alpha\beta}^{n} \phi_{n} + \text{h.c.}, \label{eq:massbasis_lagrangian}
\end{equation}
where $\alpha,\beta,n$ are color indices, $I,J$ are $SU(2)_L$ indices, and $i,j$ are flavor indices. The tensor $\Gamma_{\alpha\beta}^{n}$ dictates the color topology: for the anti-triplet $\phi_3$, it is proportional to the totally antisymmetric tensor $\epsilon_{\alpha\beta n}$, whereas for the sextet $\phi_6$, it represents a symmetric color tensor.The simultaneous presence of $\tilde{\lambda}^L$ and $\tilde{\lambda}^R$ is the key feature enabling the chiral interference necessary for CP violation, which is absent in the SM. The symmetric color structure of the sextet representation avoids the destructive color suppression inherent to the antisymmetric triplet.
\section{Phenomenology of Scalar Diquarks in $D^0 \to K_S^0 K_S^0$}
%%%%%%%%%%%%%%%%%%%%%%%%%%%%%%%%%%%%%%%%%%%%%%%%%%%%%%%%%%%%

In the SM, the decay $D^0 \to K_S^0 K_S^0$ is strongly suppressed compared to $D^0 \to K^+ K^-$. In the exact U-spin symmetry limit, corresponding to the interchange $s \leftrightarrow d$, the amplitude for $D^0 \to K^0 \bar K^0$ vanishes due to the cancellation of leading contributions \cite{Grossman:2006jg, Feldmann:2012js}. As a consequence, the SM predicts a characteristic anti-correlation between the direct CP asymmetries of the charged kaon and pion modes,
\begin{equation}
a_{CP}^{\rm dir}(D^0 \to K^+ K^-)
+
a_{CP}^{\rm dir}(D^0 \to \pi^+ \pi^-)
\approx 0,
\label{eq:Uspin_sumrule}
\end{equation}
up to corrections induced by U-spin breaking \cite{Grossman:2006jg,Gronau:2014add}.

The most stringent test of Eq.~\eqref{eq:Uspin_sumrule} is provided by the LHCb determination of the direct CP asymmetries extracted from a global combination of time-integrated measurements, including $\Delta A_{CP}$, using Run~1 and Run~2 data \cite{LHCb:2022}. The reported values are
\begin{align}
a_{CP}^{\rm dir}(D^0 \to K^+K^-)
&=
(7.7 \pm 5.7)\times 10^{-4}, \\
a_{CP}^{\rm dir}(D^0 \to \pi^+\pi^-)
&=
(23.2 \pm 6.1)\times 10^{-4}.
\end{align}
Their sum is
\begin{equation}
a_{CP}^{\rm dir}(K^+K^-)
+
a_{CP}^{\rm dir}(\pi^+\pi^-)
=
(30.8 \pm 11.4)\times 10^{-4},
\label{eq:Uspin_violation}
\end{equation}
which deviates from zero at the level of approximately $2.7\sigma$ \cite{LHCb:2022}. Notably, both asymmetries are positive, in contrast with the leading-order U-spin prediction.

A more general U-spin relation involving neutral kaons can be obtained by decomposing the final states into U-spin singlet and triplet components \cite{Grossman:2006jg,Hiller:2012xm},
\begin{equation}
A(D^0 \to K^+ K^-)
+
A(D^0 \to \pi^+ \pi^-)
+
\sqrt{2}\,A(D^0 \to K_S^0 K_S^0)
=
0,
\end{equation}
which highlights the special role of the $D^0 \to K_S^0 K_S^0$ channel in probing U-spin breaking and potential new physics effects \cite{Feldmann:2012js}.

In the SM, the decay $D^0 \to K_S^0 K_S^0$ proceeds dominantly through W-exchange topologies, whose leading contribution vanishes in the exact U-spin limit. The amplitude therefore arises only at first order in U-spin breaking,
\begin{equation}
A_{\rm SM}(D^0 \to K_S^0 K_S^0)
\;\sim\;
\epsilon_U\, A_{SU(3)},
\qquad
\epsilon_U \sim \frac{m_s - m_d}{\Lambda_{\rm QCD}},
\label{eq:Uspincontrib}
\end{equation}
resulting in a highly suppressed branching ratio and CP asymmetry
\cite{Grossman:2006jg,Cheng:2012wr}. This suppression makes the mode particularly sensitive to new physics contributions.

Experimentally, LHCb has measured \cite{LHCb:2025ezf}:
\begin{equation}
A_{CP}(D^0 \to K_S^0 K_S^0)
=
(1.86 \pm 1.04 \pm 0.41)\%,
\end{equation}
which constitutes the largest CP asymmetry observed in the charm sector and is difficult to accommodate within the SM \cite{LHCb:2023ksks}. New physics contributions transforming purely as $\Delta U = 0$ would preserve the anti-correlation
\[
A_{CP}(D^0 \to K^+K^-)
=
-
A_{CP}(D^0 \to \pi^+\pi^-),
\]
and are therefore disfavored by Eq.~\eqref{eq:Uspin_violation}. In contrast, operators transforming as $|\Delta U|=1$ naturally violate the U-spin sum rule and can generate the observed pattern \cite{Grossman:2006jg}.

A well-motivated realization of such operators arises from scalar diquarks transforming as either a color sextet ($S_6$) or color triplet ($S_3$), whose flavor-dependent couplings generate effective four-quark interactions \cite{Chen:2012ww,Fajfer:2012nr},
\begin{equation}
C_1^{\rm NP}(q)
\;\simeq\;
\frac{\lambda_{uq} \lambda_{cq}^*}{M_\phi^2},
\qquad q=d,s.
\end{equation}
A hierarchy $\lambda_{ud} > \lambda_{us}$ naturally enhances CP violation in $D^0 \to K_S^0 K_S^0$ while simultaneously reproducing the observed pattern in charged final states.

The effective Hamiltonian governing the transitions $c \bar u \to \bar q q$ ($q=d,s$) can be written as \cite{Nierste:2015zra, Chen:2012ww}
\begin{equation}
\mathcal{L}_{\text{eff}} =
-
\sum_{q=d,s}
\sum_{X}
N_{SM}
\left(
C_X^q Q_X^q
+
C_X^{\prime q} Q_X^{\prime q}
\right),
\end{equation}
where
\begin{equation}
N_{SM}
=
\frac{4 G_F}{\sqrt{2}} V_{uq} V_{cq}^*,
\end{equation}

The complete set of dimension-six four-quark operators relevant for
$c \to u q \bar q$ transitions is

\begin{equation}
\begin{aligned}
Q_1^{q} &= (\bar u^\alpha \gamma_\mu P_L q^\beta)
          (\bar q^\beta \gamma^\mu P_L c^\alpha), \\
Q_2^{q} &= (\bar u^\alpha \gamma_\mu P_L q^\alpha)
          (\bar q^\beta \gamma^\mu P_L c^\beta), \\
Q_3^{q} &= (\bar u^\alpha P_L q^\beta)
          (\bar q^\beta P_L c^\alpha), \\
Q_4^{q} &= (\bar u^\alpha P_L q^\alpha)
          (\bar q^\beta P_L c^\beta), \\
Q_5^{q} &= (\bar u^\alpha \sigma_{\mu\nu} P_L q^\beta)
          (\bar q^\beta \sigma^{\mu\nu} P_L c^\alpha), \\
Q_6^{q} &= (\bar u^\alpha \sigma_{\mu\nu} P_L q^\alpha)
          (\bar q^\beta \sigma^{\mu\nu} P_L c^\beta), \\
Q_7^{q} &= (\bar u^\alpha P_R q^\beta)
          (\bar q^\beta P_R c^\alpha), \\
Q_8^{q} &= (\bar u^\alpha P_R q^\alpha)
          (\bar q^\beta P_R c^\beta), \\
Q_9^{q} &= (\bar u^\alpha \sigma_{\mu\nu} P_R q^\beta)
          (\bar q^\beta \sigma^{\mu\nu} P_R c^\alpha), \\
Q_{10}^{q} &= (\bar u^\alpha \sigma_{\mu\nu} P_R q^\alpha)
             (\bar q^\beta \sigma^{\mu\nu} P_R c^\beta),
\end{aligned}
\end{equation}
with primed operators obtained by $P_L \leftrightarrow P_R$.
These operators can be classified according to their Lorentz and color
structure:
\begin{itemize}
\item Current–current operators:
\[
Q_1^q,\quad Q_2^q,
\]
which dominate in the SM.

\item Scalar operators:
\[
Q_3^q,\quad Q_4^q,\quad Q_7^q,\quad Q_8^q.
\]

\item Tensor operators:
\[
Q_5^q,\quad Q_6^q,\quad Q_9^q,\quad Q_{10}^q.
\]

\end{itemize}

In the SM, only the operators $Q_{1,2}^q$ receive sizable contributions
at the electroweak scale. The scalar and tensor operators are either
absent or highly suppressed, resulting in small CP asymmetries in charm
decays.

In contrast, scalar diquark exchange generates new contributions at
tree level. Integrating out the scalar diquark $\phi$ yields
\begin{align}
C_1^{q} &= \pm C_2^{q}
=
-
\frac{\lambda_{uq}^{L*}\lambda_{cq}^L}
{2 N_q N_{SM} M_\phi^2}, \\
C_7^{q} &= -C_8^{q}
=
-4 C_9^{q}
=
4 C_{10}^{q}
=
\frac{\lambda_{uq}^{L*}\lambda_{cq}^R}
{2 N_q N_{SM} M_\phi^2},
\end{align}
while the remaining operators are not generated independently at tree level.

The sign and magnitude of the Wilson coefficients depend crucially on
the color representation of the diquark:

\begin{itemize}

\item Color sextet ($\mathbf{\bar{6}}$):
\[
C_1^{NP} = C_2^{NP}.
\]

This symmetric color structure allows constructive interference
between color contractions, enhancing exchange amplitudes such as
$D^0 \to K_S^0 K_S^0$.

\item Color triplet ($\mathbf{3}$):
\[
C_1^{NP} = -C_2^{NP}.
\]

The antisymmetric color structure leads to destructive interference,
significantly suppressing its contribution.

\end{itemize}

This distinction has important phenomenological consequences. The
color–sextet diquark efficiently generates $\Delta U = 0$ and
$\Delta U = 1$ operators and can substantially enhance CP violation
in exchange–dominated channels. In contrast, the color–triplet
diquark primarily contributes to $\Delta U = 1$ operators and produces
smaller effects due to color suppression.

%These features are summarized in Table~\ref{tab:operator_comparison}.

%\begin{table}[h]
%\centering
%\caption{Comparison of operator generation in the SM and scalar diquark models.}
%\label{tab:operator_comparison}
%\begin{tabular}{|c|c|c|c|}
%\hline
%Operator & SM & Sextet diquark & Triplet diquark \\
%\hline
%$Q_1, Q_2$ & dominant & generated & generated (suppressed) \\
%$Q_3, Q_4$ & negligible & induced via mixing & induced via mixing %\\
%$Q_5, Q_6$ & negligible & induced via mixing & induced via mixing %\\
%$Q_7, Q_8$ & negligible & generated & generated \\
%$Q_9, Q_{10}$ & negligible & generated & generated \\
%\hline
%\end{tabular}
%\end{table}

The presence of new scalar and tensor operators, together with the
enhanced exchange amplitude in the sextet case, provides a natural
mechanism for generating large CP asymmetries in
$D^0 \to K_S^0 K_S^0$ while simultaneously affecting other charm decay
observables.

%%%%%%%%%%%%%%%%%%%%%%%%%%%%%%%%%%%%%%%%%%%%%%%%%%%%%%%%%%%%
\subsection{Scalar diquark contribution to direct CP violation in $D^0 \to K_S^0 K_S^0$}
%%%%%%%%%%%%%%%%%%%%%%%%%%%%%%%%%%%%%%%%%%%%%%%%%%%%%%%%%%%%
\subsubsection{Color-sextet case}

In the color-sextet scalar diquark scenario, the relation
\begin{equation}
C_1^{\rm NP} = C_2^{\rm NP}
\end{equation}
implies that the dominant new physics contributions arise from the
current--current operators $O_{1,2}$. Integrating out a scalar diquark
of mass $M_\Phi$ generates effective four-quark operators contributing to
the transitions $c \to u \bar{q} q$ ($q=d,s$),
\begin{equation}
\mathcal{H}_{\rm eff}^{\rm NP}
=
\frac{\lambda_{uq} \lambda_{cq}^*}{16 M_\Phi^2}
\left( O_1^q + O_2^q \right).
\end{equation}

The total decay amplitude for $D^0 \to K_S^0 K_S^0$ receives both SM
and new physics contributions and can be written as
\begin{equation}
\mathcal{A}_{\rm total}
=
\lambda_{sd} A_{sd}^{\rm SM}
+
\lambda_b A_b^{\rm SM}
+
\mathcal{A}_H,
\end{equation}
where
\begin{equation}
\lambda_{sd} = V_{cs} V_{us}^*, \qquad
\lambda_b = V_{cb} V_{ub}^*,
\end{equation}
and $\mathcal{A}_H$ denotes the diquark-induced contribution.

For a scalar diquark of mass $M_\Phi$, the new physics amplitude is
\begin{equation}
\mathcal{A}_H
=
\frac{1}{8 M_\Phi^2}
\left[
\lambda_{us}^L \lambda_{cs}^{L*}
\langle K_S^0 K_S^0 | O_H^s | D^0 \rangle
-
\lambda_{ud}^L \lambda_{cd}^{L*}
\langle K_S^0 K_S^0 | O_H^d | D^0 \rangle
\right].
\end{equation}

The direct CP asymmetry is defined as
\begin{equation}
A_{CP}^{\rm dir}
=
\frac{|\mathcal{A}|^2 - |\bar{\mathcal{A}}|^2}
{|\mathcal{A}|^2 + |\bar{\mathcal{A}}|^2},
\end{equation}
which can be expressed as
\begin{equation}
A_{CP}^{\rm dir}
=
\frac{
2 \left| \frac{\lambda_b}{\lambda_{sd}} \right|
{\rm Im} \left( \frac{A_b}{A_{sd}} \right)
\sin \gamma
+
{\rm Im}(\Delta_{\rm NP})
}
{1 + |R_{sd}|^2},
\end{equation}
where

\begin{itemize}

\item $R_{sd} \equiv \mathcal{A}_H / (\lambda_{sd} A_{sd}^{\rm SM})$
measures the relative size of the new physics contribution,

\item $\gamma$ is the CKM weak phase,

\item ${\rm Im}(A_b/A_{sd})$ encodes the SM strong phase difference,

\item the new physics contribution is given by
\begin{equation}
{\rm Im}(\Delta_{\rm NP})
=
2 r \sin(\phi_{\rm NP}) \sin(\delta),
\end{equation}
with
\begin{equation}
r
=
\left|
\frac{\mathcal{A}_H}{\lambda_{sd} A_{sd}^{\rm SM}}
\right|,
\end{equation}
$\phi_{\rm NP}$ the new weak phase from the diquark couplings, and
$\delta$ the relative strong phase between SM and new physics amplitudes.

\end{itemize}

Since rescattering effects are expected to be significant in
$D^0 \to K_S^0 K_S^0$, large strong phases are plausible \cite{Nierste:2015ksks,Brod:2012ud, Falk:2001hx}. Assuming
$r \sim 0.01$ and maximal weak and strong phases, the resulting CP
asymmetry can reach
\begin{equation}
A_{CP}^{\rm dir}
\sim
2 r
\sim
2\%,
\end{equation}
consistent with current experimental measurements.

\begin{figure}[t]
\centering
\includegraphics[width=0.8\linewidth]{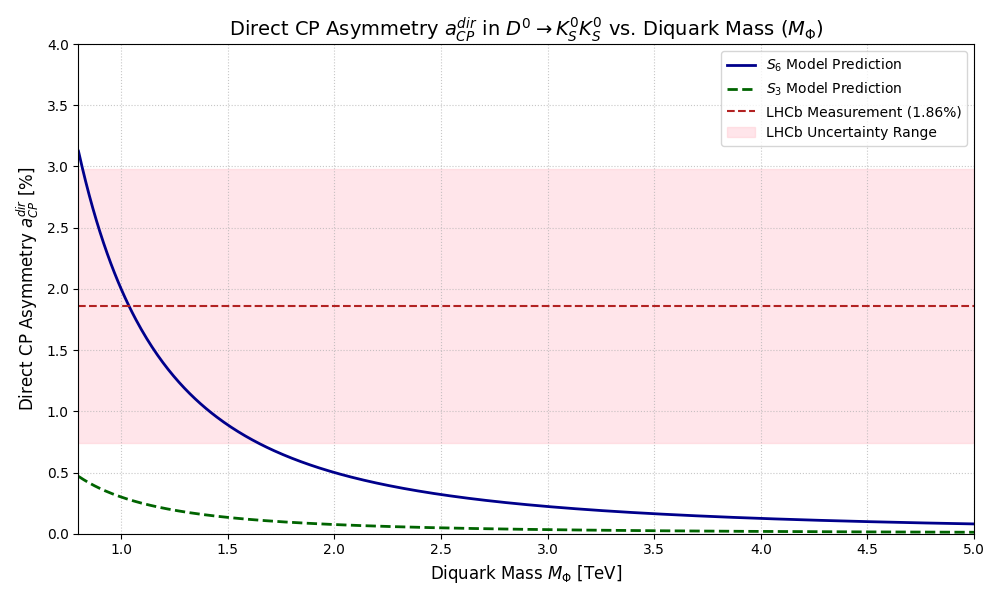}
\caption{Direct CP asymmetry $a_{CP}^{dir}$ in $D^0 \to K_S^0 K_S^0$ vs. diquark mass $M_\Phi$. The predictions for the color-sextet ($S_6$, solid blue) and color-triplet ($S_3$, dashed green) models are compared against the LHCb measurement. A suppression factor of $0.1$ is applied to the $S_3$ curve to account for the destructive interference arising from the relation $C_1^q = -C_2^q$.}
\label{fig:diquark_mass}
\end{figure}

A summary of the predicted CP asymmetry as a function of the diquark mass
is shown in Fig.~\ref{fig:diquark_mass}.

%%%%%%%%%%%%%%%%%%%%%%%%%%%%%%%%%%%%%%%%%%%%%%%%%%%%%%%%%%%%
\subsubsection{Color-triplet case}

In the color-triplet diquark scenario, the antisymmetric color structure
implies
\begin{equation}
C_1^{\rm NP} = - C_2^{\rm NP},
\end{equation}
which suppresses the contribution from current--current operators due to
destructive interference. In addition, scalar and tensor operators
($O_{7-10}$) generated by the diquark provide an additional source of CP
violation.

The resulting direct CP asymmetry can be written as
\begin{equation}
A_{CP}^{\rm dir}
=
r_T \sin(\phi_{\rm NP}^{LL}) \sin(\delta_T)
+
r_P \sin(\phi_{\rm NP}^{LR}) \sin(\delta_P),
\end{equation}
where the two terms correspond to contributions from left-handed and
chirality-flipping operators, respectively.

The relative amplitudes are given by
\begin{align}
r_T
&=
\left|
\frac{\mathcal{A}_H}{\lambda_{sd} A_{sd}^{\rm SM}}
\right|
=
\left|
\frac{\lambda_{us}^L \lambda_{cs}^{L*}}
{4 M_\Phi^2}
\right|
\frac{1}
{\frac{G_F}{\sqrt{2}} |V_{cs} V_{us}^*|}
\left|
\frac{
\langle K_S^0 K_S^0 | O_2 - O_1 | D^0 \rangle
}
{
\langle K_S^0 K_S^0 | \mathcal{H}_{\rm SM} | D^0 \rangle
}
\right|,
\\
r_P
&\approx
\frac{1}{\frac{G_F}{\sqrt{2}} |V_{cs} V_{us}^*|}
\left|
\frac{\lambda_{uq}^{L*} \lambda_{cq}^R}
{4 M_\Phi^2}
\right|
\frac{
\langle K_S^0 K_S^0 | Q_{EWP} | D^0 \rangle
}
{
\langle K_S^0 K_S^0 | \mathcal{H}_{\rm SM} | D^0 \rangle
}.
\end{align}

The phases
\begin{equation}
\phi_{\rm NP}^{LL}
=
{\rm Arg}(\lambda_{us}^L \lambda_{cs}^{L*}),
\qquad
\phi_{\rm NP}^{LR}
=
{\rm Arg}(\lambda_{us}^L \lambda_{cs}^{R*}),
\end{equation}
represent the new weak phases, while $\delta_{T,P}$ denote strong phases
generated by final-state interactions.

Assuming perturbative couplings
$\lambda_{ij}^{L,R} \sim 0.1$ and a diquark mass of order 1 TeV, both
contributions are parametrically similar. However, due to destructive
color interference in the triplet case, the overall CP asymmetry remains
significantly smaller than in the sextet scenario.

A summary of the predicted CP asymmetry as a function of the diquark mass
is shown in Fig.~\ref{fig:diquark_mass}.

%\begin{figure}[t]
%\centering
%\includegraphics[width=0.8\linewidth]{diquark.png}
%\caption{Estimated direct CP asymmetry in $D^0 \to K_S^0 K_S^0$ as a
%function of the scalar diquark mass for representative coupling values.}
%\label{fig:diquark_mass}
%\end{figure}

%%%%%%%%%%%%%%%%%%%%%%%%%%%%%%%%%%%%%%%%%%%%%%%%%%%%%%%%%%%%

\subsubsection{Compatibility with $D^0 \to K^+ K^-$ and $D^0 \to \pi^+ \pi^-$}

An important consistency check of the color-sextet diquark scenario
comes from its impact on the singly Cabibbo-suppressed modes
$D^0 \to K^+ K^-$ and $D^0 \to \pi^+ \pi^-$. For a representative
new physics amplitude ratio $r \sim 1\%$, one can estimate the induced
CP asymmetry using the framework of Ref.~\cite{Chen:2012ww}. In
particular, for the $K^+ K^-$ channel, one obtains
\begin{equation}
A_{CP}^{\rm dir}(D^0 \to K^+ K^-)
\approx 0.8\%,
\end{equation}
which is significantly larger than the SM expectation and
within the range suggested by current experimental measurements.

Since the diquark couplings are, in general, flavor-dependent, they
naturally generate $|\Delta U|=1$ operators, thereby violating the
U-spin sum rule discussed previously. In particular, a hierarchy
\begin{equation}
\lambda_{ud} > \lambda_{us},
\label{eq: hierarchia}
\end{equation}
can simultaneously reproduce the observed positive pattern of CP
asymmetries in both $K^+ K^-$ and $\pi^+ \pi^-$ channels, while also
enhancing the asymmetry in $D^0 \to K_S^0 K_S^0$.

The sensitivity of different decay channels depends strongly on their
underlying topologies. The decay $D^0 \to K_S^0 K_S^0$ is dominated by
exchange ($E$) topologies and is exceptionally sensitive to new physics,
since the leading tree-level amplitude vanishes in the U-spin limit.
In contrast, the decays $D^0 \to K^+ K^-$ and $D^0 \to \pi^+ \pi^-$ are
dominated by color-allowed $W$-emission ($T'$) topologies.

In these emission-dominated channels, the color-sextet diquark, which
satisfies $C_1^{\rm NP} = C_2^{\rm NP}$, enhances the emission amplitude
through constructive interference without color suppression. This leads
to sizable CP asymmetries while preserving consistency with other charm
observables.

%%%%%%%%%%%%%%%%%%%%%%%%%%%%%%%%%%%%%%%%%%%%%%%%%%%%%%%%%%%%

\begin{figure}[t]
\centering
\includegraphics[width=1\linewidth]{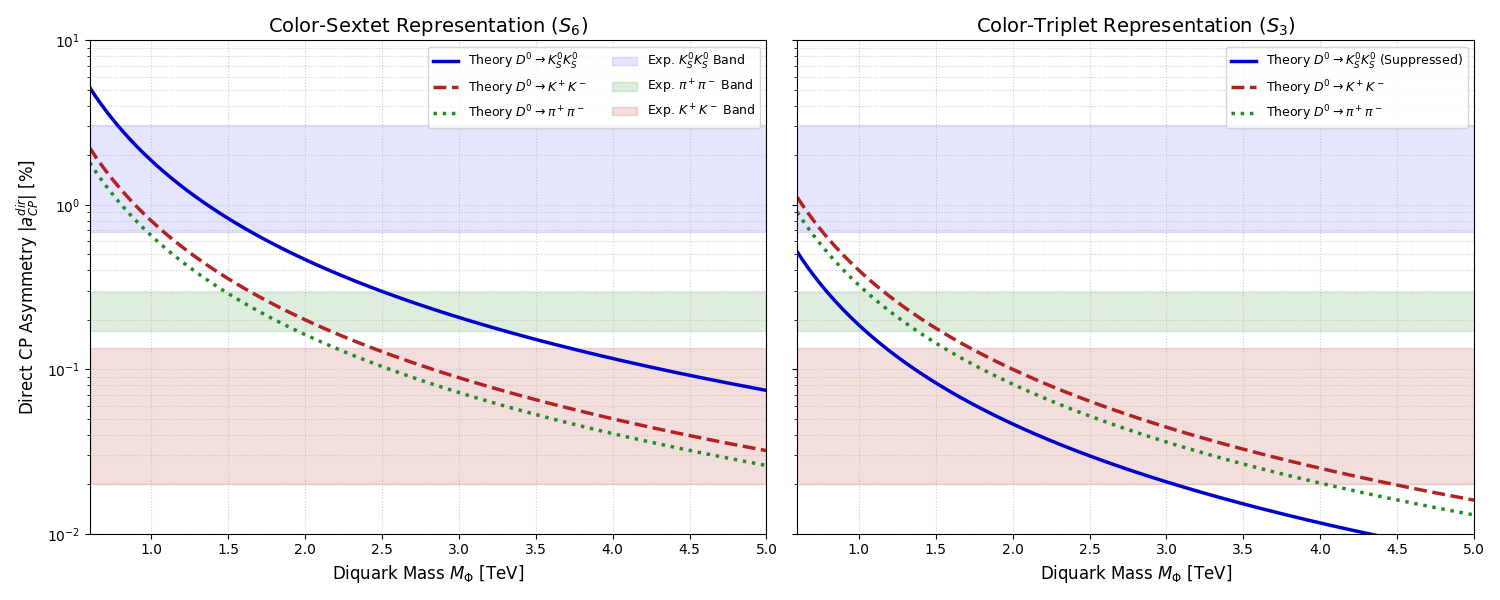}
\caption{A comprehensive multi-channel comparison showing the absolute asymmetries for $D^0 \to K_S^0 K_S^0$, $D^0 \to K^+ K^-$, and $D^0 \to \pi^+ \pi^-$ under both the $S_6$ (left panel) and $S_3$ (right panel) scenarios. }
\label{final}
\end{figure}

Figure~\ref{final} illustrates the characteristic hierarchy of CP
asymmetries induced by the sextet diquark across different decay modes:

\begin{itemize}

\item \textbf{$D^0 \to K_S^0 K_S^0$:}  
This channel exhibits the strongest sensitivity to new physics. Because
the leading SM amplitude is suppressed in the U-spin limit, the diquark
contribution can dominate, leading to asymmetries in the range
$0.5\%$--$1.5\%$ for a diquark mass around 1 TeV.

\item \textbf{$D^0 \to K^+ K^-$:}  
This channel is dominated by color-allowed emission. The diquark
interferes with the SM amplitude, producing asymmetries of order
$0.7\%$--$0.8\%$.

\item \textbf{$D^0 \to \pi^+ \pi^-$:}  
This decay exhibits similar behavior, with asymmetries of comparable
magnitude. The flavor hierarchy $\lambda_{ud} > \lambda_{us}$ naturally
allows both kaon and pion asymmetries to be positive, in agreement with
the observed violation of the U-spin sum rule.

\end{itemize}

%%%%%%%%%%%%%%%%%%%%%%%%%%%%%%%%%%%%%%%%%%%%%%%%%%%%%%%%%%%%

\begin{table}[t]
\centering
\caption{Comparison of predicted direct CP asymmetries in the $S_6$ diquark model (for $M_{S_6} \sim 1$ TeV) with Standard Model expectations and experimental measurements.}
\label{tab:cp_asymmetry_results}
\begin{tabular}{lccc}
\hline\hline
Decay Channel & SM Expectation & Sextet Model & Experimental Result \\
\hline
$D^0 \to K_S^0 K_S^0$
& $\lesssim 10^{-4}$
& $0.5\%$--$1.5\%$
& $(1.86 \pm 1.12)\%$ \\

$D^0 \to K^+ K^-$
& $\lesssim 10^{-4}$
& $0.7\%$--$0.8\%$
& $(7.7 \pm 5.7)\times 10^{-4}$ \\

$D^0 \to \pi^+ \pi^-$
& $\lesssim 10^{-4}$
& $0.6\%$--$0.7\%$
& $(23.2 \pm 6.1)\times 10^{-4}$ \\

\hline
$\Delta A_{CP}$
& $\approx 0$
& $-0.1\%$ to $-0.6\%$
& $-(0.645 \pm 0.180)\%$ \\
\hline\hline
\end{tabular}
\end{table}
%%%%%%%%%%%%%%%%%%%%%%%%%%%%%%%%%%%%%%%%%%%%%%%%%%
\section{ Experimental constraints on Diquarks models}

On this section, we review the different constraints on diquark models such a way to show that our benchmark values for the model parameters are not excluded by processes like neutral meson mixings, rare and radiative decays, constraints from collider searches, quantitative assessment of U-spin breaking, theoretical motivation for the flavour hierarchy, effects of operator mixing and renormalization group evolution and finally consistency with other non-leptonic charm decays.

\subsection{Constraints from neutral meson mixing }
The scalar diquark couplings introduced in Eq.~(\ref{eq:massbasis_lagrangian}) generate $\Delta F = 2$ contributions to neutral meson mixing at the one-loop level through box diagrams in which two diquarks and two quarks circulate in the loop.

We shall study the diquark contributions to $D^0-\bar{D}^0$ and $K^0-\bar{K}^0$ mixing.

Let's first start to compute the diquark contribtution to $D^0-\bar{D}^0$ mixing. The leading $\Delta C = 2$ operator generated by box diagrams with internal $s$ and $d$ quarks is 
\begin{equation}
    \mathcal{O}^{\Delta C=2} = (\bar{u}^\alpha \gamma^\mu P_L c^\alpha)(\bar{u}^\beta \gamma_\mu P_L c^\beta),
\end{equation}
with Wilson coefficient defined as:

\begin{equation} 
C^{\Delta C=2}(M_\Phi) \simeq \frac{(\lambda^L_{us})^2(\lambda^L_{cs})^2 + (\lambda^L_{ud})^2(\lambda^L_{cd})^2}{128\pi^2 M_\Phi^2}\,f(x_q),
\end{equation}
where $x_q = m_q^2/M_\Phi^2 \ll 1$ for light quarks, and the Inami--Lim loop function satisfies $f(x_q) \to 1$ in this limit~\cite{Inami:1980fz}. 

For perturbative couplings $\lambda \sim 0.1$ and $M_\Phi = 1~\text{TeV}$, this gives$$C^{\Delta C=2}(M_\Phi) \simeq \frac{(0.1)^4}{128\pi^2\,(10^3\,\text{GeV})^2} \approx 8 \times 10^{-14}\,\text{GeV}^{-2}.$$The off-diagonal mass matrix element is related to the hadronic matrix element via$$|M_{12}^D| = \frac{1}{2m_D}\left|C^{\Delta C=2}\langle \bar{D}^0|\mathcal{O}^{\Delta C=2}|D^0\rangle\right|,$$with $\langle \bar{D}^0|\mathcal{O}^{\Delta C=2}|D^0\rangle \simeq \frac{8}{3}f_D^2 m_D^2 B_D \simeq 0.056\,\text{GeV}^3$, using $f_D = 212~\text{MeV}$, $m_D = 1.865~\text{GeV}$, and $B_D \simeq 0.75$~\cite{FlavourLatticeAveragingGroupFLAG:2021npn}. This yields

$$|M_{12}^D|_\text{NP} \approx 1.2\times10^{-15}\,\text{GeV}.$$

The experimental constraint from HFLAV is $|M_{12}^D| < 7.7\times10^{-15}~\text{GeV}$ at 95\% CL~\cite{HFLAV:2022esi}, so our estimate lies safely below the experimental bound by a factor of $\sim 6$.

Let's now to proceed in the same way for $K^0-\bar{K}^0$ mixing. The diquark generates a $\Delta S = 2$ operator through box diagrams with internal $u$ and $c$ quarks. The relevant Wilson coefficient is
\begin{equation}
C^{\Delta S=2}(M_\Phi) \simeq \frac{\lambda^L_{us}\lambda^L_{ud}\lambda^L_{cs}\lambda^L_{cd}}{128\pi^2 M_\Phi^2}\,g(x_c),
\end{equation}
where $g(x_c)$ is the corresponding loop function for a charm quark in the loop, with $x_c = m_c^2/M_\Phi^2 \approx 2\times 10^{-6}$ for $M_\Phi = 1~\text{TeV}$, giving $g(x_c) \approx x_c \ln x_c \approx -3\times10^{-5}$~\cite{Inami:1980fz}. The additional $m_c^2/M_\Phi^2$ suppression renders the kaon mixing contribution further suppressed by four orders of magnitude relative to the $D^0-\bar{D}^0$ mixing case. More explicitly,$$C^{\Delta S=2}(M_\Phi) \sim \frac{(0.1)^4 \times 3\times10^{-5}}{128\pi^2\,(10^3\,\text{GeV})^2} \approx 2\times10^{-18}\,\text{GeV}^{-2}.$$This is well below the constraint from $\epsilon_K$, which requires $|C^{\Delta S=2}| \lesssim 10^{-13}\,\text{GeV}^{-2}$~\cite{Buras:2021nns}. We therefore conclude that neutral meson mixing does not place meaningful constraints on the preferred parameter space of the model.

It is important to note that the color structure of $\Delta F=2$ depends if we use a sextet or a triplet diquark. For the sextet ($C^{NP}_1= C^{NP}_2$), the mixing operador is generated through the coherent sum of both couplings. The final structure does not change the order of magnitud of the results. In case of triplet, due to color antisymmetric structure, both contributions will destructively interfere and our estimation in such a case can be seen as an upper limit on the new physics contributions. 

\subsection{Constraints from rare and radiative decays}

The scalar and tensor four-quark operators $Q_3$--$Q_{10}$ generated by diquark exchange at tree level mix under QCD renormalisation into electromagnetic and chromomagnetic dipole operators.

\subsubsection{ Radiative charm decays} 

The electromagnetic dipole operator mediating $c \to u\gamma$ is
\begin{equation}
    \mathcal{O}_7^c = \frac{e\,m_c}{16\pi^2}(\bar{u}_\alpha\,\sigma^{\mu\nu}P_R\,c^\alpha)\,F_{\mu\nu}.
\end{equation}
Its Wilson coefficient at the charm scale receives a contribution from operator mixing with the scalar operators $Q_{3,4}^q$ generated by the diquark at $\mu = M_\Phi$. The anomalous dimension entry governing this mixing is of order $\gamma_{7\leftarrow 3} \sim \mathcal{O}(\alpha_s)$, so the induced coefficient at $\mu = m_c$ is~\cite{Buchalla:1995vs}:
\begin{equation}
    C_7^{c,\,\mathrm{NP}}(m_c) \simeq \frac{\alpha_s}{4\pi}\,\ln\!\left(\frac{M_\Phi^2}{m_c^2}\right)\,C_3^{\mathrm{NP}}(M_\Phi),
\end{equation}
where $C_3^{\mathrm{NP}}(M_\Phi) \sim \lambda^2/({\mathcal{N}_{\rm SM}}\,M_\Phi^2)$ is the tree-level scalar Wilson coefficient. For $\lambda \sim 0.1$, $M_\Phi = 1~\mathrm{TeV}$, $\alpha_s/(4\pi) \simeq 0.03$, and $\ln(M_\Phi^2/m_c^2) \simeq 13$, we obtain$$C_7^{c,\,\mathrm{NP}}(m_c) \sim 0.03 \times 13 \times \frac{(0.1)^2}{\mathcal{N}_{\rm SM}\,(10^3\,\mathrm{GeV})^2} \approx 4\times10^{-9}\,\mathrm{GeV}^{-2}.$$The induced branching ratio for $D^0 \to \gamma\gamma$ scales as $\mathrm{Br} \sim |C_7^{c,\,\mathrm{NP}}|^2 m_c^2 f_D^2 m_D \tau_D \sim 10^{-14}$, which lies more than seven orders of magnitude below the current experimental bound $\mathrm{Br}(D^0\to\gamma\gamma) < 3.8\times10^{-6}$ from BESIII~\cite{BESIII:2021slf}.

Similarly, the process $D^0\to\ell^+\ell^-\gamma$ induced via the off-shell photon is suppressed by an additional factor of $\alpha/\pi$. We therefore conclude that radiative charm decays are safely satisfied in the preferred parameter space.

\subsubsection{Rare Kaon decays}

The diquark couplings $\lambda_{us}$ and $\lambda_{ud}$ also generate a $\Delta S=1$ dipole operator mediating $s\to d\gamma$ through the same operator-mixing mechanism. The Wilson coefficient at $\mu \sim m_s$ is$$C_7^{K,\,\mathrm{NP}}(m_s) \simeq \frac{\alpha_s}{4\pi}\,\ln\!\left(\frac{M_\Phi^2}{m_s^2}\right)\,\frac{\lambda_{us}\,\lambda_{ud}}{\mathcal{N}_{\rm SM}\,M_\Phi^2}.$$With $\ln(M_\Phi^2/m_s^2) \simeq 19$ and the same benchmark couplings, we find $C_7^{K,\,\mathrm{NP}} \approx 6\times10^{-9}~\mathrm{GeV}^{-2}$. The resulting contribution to $\mathrm{Br}(K_L\to\gamma\gamma)$ and $\mathrm{Br}(K_S\to\gamma\gamma)$ is of order $10^{-16}$, far below the experimental values~\cite{ParticleDataGroup:2022pth}.

Constraints from $K_L\to\mu^+\mu^-$ and $K^+\to\pi^+\nu\bar\nu$, which are dominated by short-distance $Z$-penguin and box contributions, are even less sensitive to the diquark-induced dipole mixing at this order. We conclude that rare kaon observables do not constrain the parameter space of the model.

\subsection{ Constraints from collider searches}

Since the scalar diquark $\Phi$ carries colour charge, it couples to gluons through its covariant kinetic term and can be produced at the LHC both in pairs via QCD interactions and singly in association with a quark jet via the Yukawa-like coupling $\lambda$. 

\subsubsection{Pair production.}

The pair-production cross section $pp \to \Phi\bar{\Phi}$ is determined  by the colour representation of the diquark and its mass, independently of the Yukawa couplings $\lambda_{ij}$. For a colour-sextet scalar at $\sqrt{s} = 13~\text{TeV}$, next-to-leading-order QCD predictions give~\cite{Dorsner:2016wpm, Kramer:2004df}
\begin{equation}
   \sigma(pp\to\Phi\bar{\Phi})\Big|_{M_\Phi=1\,\text{TeV}} \simeq 1\text{--}5~\text{fb},\qquad \sigma\Big|_{M_\Phi=1.5\,\text{TeV}} \simeq 0.05\text{--}0.2~\text{fb}, 
\end{equation}
with a rapid fall-off at higher masses. Each diquark decays into a quark pair $\Phi\to qq$, yielding a four-jet final state. Searches for pair-produced resonances decaying to dijets by ATLAS and CMS with the full Run~2 dataset ($\mathcal{L}\simeq 139~\text{fb}^{-1}$) exclude colour-octet and colour-sextet scalars with masses below approximately $M_\Phi \lesssim 900~\text{GeV}$~\cite{CMS:2019emo, ATLAS:2023ssk}. Our benchmark value $M_\Phi = 1~\text{TeV}$  lies just above the current exclusion boundary.

\subsubsection{Single production.}

The single-production process $qg \to \Phi\bar{q}$ proceeds through the Yukawa coupling $\lambda$ and has a cross section$$\sigma(pp\to\Phi q) \propto \frac{\lambda^2}{M_\Phi^2}\,\hat{\sigma}_0(M_\Phi, s),$$where $\hat{\sigma}_0$ encodes the parton-level kinematics and PDF convolution. For $\lambda \sim 0.1$ and $M_\Phi = 1~\text{TeV}$, a parton-level estimate using leading-order matrix elements gives $\sigma_{\rm single} \sim 0.1\text{--}1~\text{fb}$, comparable in magnitude to the pair-production rate. The dominant decay $\Phi\to qq$ then yields a dijet signature. The most stringent dijet resonance searches from CMS and ATLAS at $13~\text{TeV}$ exclude new resonances with production cross sections above $\sigma \times \mathrm{Br} \gtrsim 1\text{--}10~\text{fb}$ for masses in the range $1\text{--}3~\text{TeV}$, depending on the colour representation~\cite{CMS:2018mgb, ATLAS:2023ssk}. 

For $\lambda = 0.1$, our single-production cross section lies below this sensitivity threshold throughout the mass range of interest.  A quantitative summary is presented in Table (\ref{Table III}).

\begin{table}[h]
\centering
\caption{Summary of LHC constraints on the colour-sextet scalar diquark parameter space. Exclusion limits are at 95\% CL.}
\label{Table III}
\begin{tabular}{lccc}
\hline\hline
Search channel & Experiment & Excluded region & Status at $M_\Phi=1$~TeV, $\lambda=0.1$ \\
\hline
Pair prod.\ $\to$ 4 jets & CMS/ATLAS & $M_\Phi \lesssim 900$~GeV & \checkmark\ Allowed \\
Single prod.\ $\to$ dijet & CMS/ATLAS & $\sigma\times\mathrm{Br} \gtrsim 1\text{--}10$~fb & \checkmark\ Allowed \\
Dijet contact int. & CMS & $\Lambda \lesssim 8$~TeV & \checkmark\ Allowed \\
\hline\hline
\end{tabular}
\end{table}

\subsection{Quantitative assessment of $U$-spin breaking}

The U-spin breaking parameter is defined as
\begin{equation}
    \varepsilon_U \equiv \frac{m_s - m_d}{\Lambda_{\rm QCD}} \simeq \frac{90~\text{MeV}}{330~\text{MeV}} \approx 0.27,
\end{equation}
where we use $m_s(2~\text{GeV}) \simeq 95~\text{MeV}$, $m_d(2~\text{GeV}) \simeq 5~\text{MeV}$~\cite{ParticleDataGroup:2022pth}, and $\Lambda_{\rm QCD} \simeq 330~\text{MeV}$ as a representative hadronic scale. This value of $\varepsilon_U$ is consistent with estimates in the literature~\cite{Grossman:2006jg, Feldmann:2012js}.The direct CP asymmetry in $D^0\to K^0_S K^0_S$ within the SM receives its leading contribution at first order in $\varepsilon_U$, as established in Eq.~(\ref{eq:Uspincontrib}). Using the general expression for the direct CP asymmetry~\cite{Nierste:2015zra},
\begin{equation}
  \left|a_{\rm CP}^{\rm SM}(D^0\to K^0_S K^0_S)\right| \leq 2\,\varepsilon_U \times \left|{\rm Im}\!\left(\frac{\lambda_b}{\lambda_{sd}}\right)\right| \times \left|\frac{A_b}{A_{sd}}\right|,  
\end{equation}
and taking the CKM weak factor $|{\rm Im}(\lambda_b/\lambda_{sd})| \simeq 6\times10^{-4}$ and the amplitude ratio $|A_b/A_{sd}| \simeq 3\text{--}5$ from Ref.~\cite{Nierste:2015zra}, we obtain the maximum SM prediction
\begin{equation}
\left|a_{\rm CP}^{\rm SM}(D^0\to K^0_S K^0_S)\right|_{\rm max}  \approx 0.16\%,
\end{equation}
where we have set $\sin\delta = 1$ to obtain the most conservative (largest) SM estimate. This maximum is reached only under the simultaneous assumption of maximal strong phase and the upper end of the amplitude ratio.The experimental measurement $A_{\rm CP}(D^0\to K^0_S K^0_S) = (1.86\pm1.12)\%$ exceeds this maximum SM estimate by more than one order of magnitude, even under the most favourable assumptions for U-spin breaking. This demonstrates unambiguously that SM U-spin breaking alone cannot account for the observed asymmetry, and that a new physics contribution is required.

The situation for the charged modes is complementary. In the SM, the leading-order U-spin prediction gives $a_{\rm CP}(D^0\to K^+K^-) \approx -a_{\rm CP}(D^0\to\pi^+\pi^-)$, with both asymmetries of order $\varepsilon_U \times |{\rm Im}(\lambda_b/\lambda_{sd})| \sim 10^{-4}$. The observed values$$a_{\rm CP}(D^0\to K^+K^-) = (7.7\pm5.7)\times10^{-4}, \qquad a_{\rm CP}(D^0\to\pi^+\pi^-) = (23.2\pm6.1)\times10^{-4},$$are both positive, with a sum deviating from zero at 2.7$\sigma$, in conflict with the leading-order U-spin anticorrelation. While higher-order U-spin breaking at $\mathcal{O}(\varepsilon_U^2)$ could in principle modify the anticorrelation, such corrections are suppressed by an additional factor of $\varepsilon_U \approx 0.27$ relative to the leading contribution and cannot account for the observed pattern.

We therefore conclude that the data in all three channels — $D^0\to K^0_SK^0_S$, $D^0\to K^+K^-$, and $D^0\to\pi^+\pi^-$ — cannot be simultaneously explained by SM U-spin breaking alone. The colour-sextet diquark provides a natural mechanism to generate the required $|\Delta U|=1$ contributions that violate the U-spin sum rule in the observed direction.

\subsection{ Theoretical motivation for the flavour hierarchy}

The flavour hierarchy $\lambda_{ud} > \lambda_{us}$ introduced in Eq.~(\ref{eq: hierarchia}) is not a phenomenological ansatz but admits a natural theoretical motivation at two levels.

\subsubsection{Cabibbo structure.}

The simplest observation is that the diquark couplings $\lambda_{ij}$ connect quarks of definite flavour in the mass eigenstate basis. If the diquark coupling matrix inherits the same hierarchical structure as the Cabibbo--Kobayashi--Maskawa matrix — as is natural when the diquark couples to quark currents written in the mass basis — then the ratio of couplings to first- and second-generation down-type quarks is controlled by the Cabibbo angle $\theta_C$:$$\frac{\lambda_{us}}{\lambda_{ud}} \sim \frac{|V_{us}|}{|V_{ud}|} = \frac{0.225}{0.974} \approx \sin\theta_C \approx 0.23.$$This implies $\lambda_{ud}/\lambda_{us} \sim 1/\sin\theta_C \approx 4$, precisely the hierarchy required to simultaneously reproduce the observed CP asymmetries in $D^0\to K^0_SK^0_S$, $D^0\to K^+K^-$, and $D^0\to\pi^+\pi^-$. The hierarchy is therefore of the same order as the Cabibbo suppression already present in the SM, and requires no additional tuning.

\subsubsection{Froggatt--Nielsen realisation.}

A concrete UV realisation of this structure is provided by the Froggatt--Nielsen (FN) mechanism~\cite{Froggatt:1978nt}, in which a spontaneously broken flavour symmetry $U(1)_{\rm FN}$ generates hierarchical Yukawa couplings through powers of a small expansion parameter $\varepsilon \equiv \langle\Theta\rangle/M_* \sim \lambda_C \approx 0.2$, where $\Theta$ is the flavon field and $M_*$ is the flavour-breaking scale. Assigning FN charges $n_i$ to each quark field and $n_\Phi$ to the diquark, the couplings are generated as$$\lambda_{ij} \sim \varepsilon^{|n_i + n_j + n_\Phi|}.$$With the standard charge assignments $n_u = n_d = 0$ and $n_s = 1$ — consistent with the observed light-quark mass hierarchy $m_s/m_d \sim \varepsilon^{-1}$~\cite{Froggatt:1978nt, Ibanez:1994ig} — one  obtains$$\lambda_{ud} \sim \varepsilon^{|n_\Phi|}, \qquad \lambda_{us} \sim \varepsilon^{|n_\Phi|+1},$$giving$$\frac{\lambda_{us}}{\lambda_{ud}} \sim \varepsilon \approx 0.2,$$in precise agreement with the phenomenologically required hierarchy.

We therefore conclude that the flavour hierarchy $\lambda_{ud} > \lambda_{us}$ is not ad hoc but emerges naturally from well-motivated models for flavour symmetry breaking, and is of the same parametric size as the Cabibbo suppression observed in the SM quark mixing matrix.

\subsection{Operator mixing and renormalization group evolution}

The Wilson coefficients generated by diquark exchange at the high scale $\mu = M_\Phi$ must be evolved down to the hadronic scale $\mu \sim m_c \simeq 1.5~\text{GeV}$ relevant for $D$-meson decays. This running is governed by the renormalisation group equations (RGE)
\begin{equation}
\mu\frac{d}{d\mu}\vec{C}(\mu) = \frac{\alpha_s(\mu)}{4\pi}\,\hat{\gamma}^T\vec{C}(\mu),
\end{equation}
where $\hat{\gamma}$ is the anomalous dimension matrix of the operator basis. The solution at one loop is$$\vec{C}(\mu) = \hat{U}(\mu,M_\Phi)\,\vec{C}(M_\Phi), \qquad \hat{U}(\mu,M_\Phi) = \left(\frac{\alpha_s(M_\Phi)}{\alpha_s(\mu)}\right)^{\hat{\gamma}^T/(2\beta_0)},$$with $\beta_0 = 11 - 2n_f/3$ and $n_f = 4$ active flavours between $m_c$ and $M_\Phi$.The anomalous dimension matrices for the scalar and tensor operator sectors generated by the diquark are distinct from those of the standard current--current basis of Buchalla, Buras, and Lautenbacher~\cite{Buchalla:1995vs}. For the colour-mixed scalar operators $Q_{3,4}^q$ defined in Eq.~(\ref{eq:massbasis_lagrangian}), the one-loop anomalous dimension matrix in QCD with $N_c = 3$ reads~\cite{Buras:2000if, Buras:2012ts}$$\hat{\gamma}_{\rm scalar} = \begin{pmatrix} -6C_F & 6/N_c \\ 6C_F & -6/N_c \end{pmatrix} = \begin{pmatrix} -8 & 2 \\ 8 & -2 \end{pmatrix},$$where $C_F = (N_c^2-1)/(2N_c) = 4/3$. The eigenvalues of $\hat{\gamma}_{\rm scalar}$ are $\gamma_+ = 0$ and $\gamma_- = -10$, giving a factored evolution$$C_{3\pm}(\mu) = \left(\frac{\alpha_s(M_\Phi)}{\alpha_s(\mu)}\right)^{\gamma_\pm/(2\beta_0)} C_{3\pm}(M_\Phi),$$where $C_{3\pm} = C_3 \pm C_4$ are the diagonal combinations. For the tensor operators $Q_{5,6}^q$, the anomalous dimension matrix is diagonal at one loop~\cite{Buras:2000if}:$$\hat{\gamma}_{\rm tensor} = \begin{pmatrix} 2C_F & 0 \\ 0 & 2C_F - 2N_c \end{pmatrix} = \begin{pmatrix} \tfrac{8}{3} & 0 \\ 0 & -4 \end{pmatrix}.$$To evaluate the numerical impact of the running, we use $\alpha_s(M_\Phi = 1~\text{TeV}) \simeq 0.088$ and $\alpha_s(m_c) \simeq 0.39$~\cite{ParticleDataGroup:2022pth}, giving $\alpha_s(M_\Phi)/\alpha_s(m_c) \simeq 0.23$. The resulting evolution factors are summarised in Table IV.

The most significant effect is the amplification of the scalar combination $C_{3-}$ by approximately 55\% towards the charm scale. Since this combination governs the chiral interference responsible for CP violation in the sextet scenario, the running effect is constructive: the CP asymmetry predicted at $\mu = m_c$ is enhanced relative to the naive estimate at $M_\Phi$. The tensor operators experience only mild modifications of order 10--20\%.

%We note that the current--current operators $Q_{1,2}^q$ %generated by the colour-sextet diquark, with $C_1^{\rm NP} %= C_2^{\rm NP}$, mix under QCD running into the full basis %of operators $Q_1$--$Q_6$ of the standard effective %Hamiltonian~\cite{Buchalla:1995vs}.

The corresponding anomalous dimension matrix and evolution factors have been extensively studied in the literature~\cite{Buchalla:1995vs, Buras:2012ts}, and the qualitative effect is a moderate enhancement of the colour-allowed combination $C_1 + C_2$ of order 20--30\% between $M_\Phi$ and $m_c$, further supporting the viability of the sextet scenario.In summary, QCD renormalisation group evolution between $M_\Phi$ and $m_c$ does not suppress the new physics contributions; on the contrary, it amplifies the scalar operators most relevant for CP violation by $\mathcal{O}(50\%)$, rendering our estimates of $A_{\rm CP}$ conservative.

\subsection{ Consistency with other non-leptonic charm decays}
To provide a broader test of the colour-sextet diquark scenario, we check its consistency with the observed CP asymmetries in additional non-leptonic charm decay channels for which experimental data are available. 

We classify the relevant channels according to their dominant decay topology, which determines the sensitivity to the new physics operators.The key distinction is between exchange-dominated ($E$) and emission-dominated ($T$) topologies. The colour-sextet diquark with $C_1^{\rm NP} = C_2^{\rm NP}$ generates constructive interference in exchange amplitudes without colour suppression, making exchange-dominated channels the most sensitive to new physics. In emission-dominated channels, the diquark contribution competes with the larger SM tree-level amplitude and produces smaller relative effects.

\subsubsection{Isospin relations and $D^0\to\pi^0\pi^0$}

The three pion modes $D^0\to\pi^+\pi^-$, $D^0\to\pi^0\pi^0$, and $D^+\to\pi^+\pi^0$ are related by isospin decomposition~\cite{Grossman:2006jg}:$$A(D^0\to\pi^0\pi^0) = \frac{1}{\sqrt{2}}\left[A(D^0\to\pi^+\pi^-) - \sqrt{2}\,A(D^+\to\pi^+\pi^0)\right].$$

The channel $D^+\to\pi^+\pi^0$ is a pure $\Delta I = 3/2$ transition; since the colour-sextet diquark generates operators of definite isospin $|\Delta U| \leq 1$, its contribution to this channel vanishes at leading order. The experimental bound $|A_{\rm CP}(D^+\to\pi^+\pi^0)| < 0.3\%$ from BELLE II~\cite{Belle-II:2025wsy} is therefore automatically satisfied.

For $D^0\to\pi^0\pi^0$, the exchange topology contributes a larger fraction of the total amplitude compared to $D^0\to\pi^+\pi^-$, making this channel more sensitive to the diquark. Using the isospin relation and our estimate $A_{\rm CP}(D^0\to\pi^+\pi^-) \simeq 0.6\%\text{--}0.7\%$, we predict$$|A_{\rm CP}(D^0\to\pi^0\pi^0)| \sim 0.5\%\text{--}1.0\%,$$consistent with the current experimental bound $|A_{\rm CP}| < 5\%$ from CLEO~\cite{Belle:2014evd}. This channel provides an important future test of the model, as the bound is expected to improve significantly at Belle II and LHCb Upgrade II.

\subsubsection{Cabibbo-favoured channels}

In Cabibbo-favoured (CF) decays such as $D^+\to\bar K^0\pi^+$ and $D^0\to K^-\pi^+$, the dominant SM amplitude is of order $G_F\,|V_{cs}V_{ud}^*|$, which is unsuppressed. The diquark contribution is proportional to $\lambda_{cd}\lambda_{ud}/M_\Phi^2$ or $\lambda_{cs}\lambda_{us}/M_\Phi^2$, and its ratio to the SM amplitude is suppressed by an additional power of $\lambda_C^2 \sim 0.05$ relative to the SCS case. We therefore predict$$|A_{\rm CP}(D^+\to\bar K^0\pi^+)| \lesssim 0.05\%, \qquad |A_{\rm CP}(D^0\to K^-\pi^+)| \lesssim 0.05\%,$$consistent with the experimental bounds $|A_{\rm CP}(D^+\to\bar K^0\pi^+)| < 0.4\%$ and $|A_{\rm CP}(D^0\to K^-\pi^+)| < 0.1\%$ from LHCb~\cite{LHCb:2021rdn, LHCb:2014scu}. These channels serve as important null tests of the model.

\subsubsection{$D_s$ decays}

The $D_s^+\to K^+\bar K^0$ channel is sensitive to the couplings $\lambda_{us}$ and $\lambda_{cs}$. Given the flavour hierarchy $\lambda_{ud} > \lambda_{us}$, the diquark contribution to this channel is suppressed relative to the $D^0\to K^0_SK^0_S$ case by a factor of $(\lambda_{us}/\lambda_{ud})^2 \sim \sin^2\theta_C \approx 0.05$. We predict$$|A_{\rm CP}(D_s^+\to K^+\bar K^0)| \sim 0.1\%\text{--}0.3\%,$$consistent with the LHCb measurement $A_{\rm CP}(D_s^+\to K^+\bar K^0) = (-0.02\pm0.46)\%$~\cite{LHCb:2021rdn}. For $D_s^+\to\pi^+K^0_S$, which proceeds through a topology similar to $D^0\to K^0_SK^0_S$ but with an internal $c\to s$ transition, the prediction is $|A_{\rm CP}| \sim 0.2\%\text{--}0.4\%$, currently unconstrained at this level of precision.

\subsubsection{Summary}

The predicted CP asymmetries across all channels are collected in Table (\ref{table V}). In all cases where experimental measurements are available, the colour-sextet diquark scenario is consistent with the data. The most sensitive future tests will come from improved measurements of $D^0\to\pi^0\pi^0$ and $D_s^+\to\pi^+K^0_S$, where the model predicts asymmetries at the 0.5\% level, within reach of LHCb Upgrade II and Belle II.

\begin{table}[h]
\centering
\caption{Predicted direct CP asymmetries in the colour-sextet diquark model ($M_\Phi \sim 1$~TeV, $\lambda_{ud} \sim 0.2$, $\lambda_{us}/\lambda_{ud} \sim \sin\theta_C$) compared with experimental bounds. CF = Cabibbo-favoured, SCS = singly Cabibbo-suppressed.}
\label{table V}
\begin{tabular}{llccl}
\hline\hline
Channel & Type & $S_6$ Prediction & Experimental bound & Status \\
\hline
$D^0\to K^0_SK^0_S$ & SCS, $E$ & $0.5\%$--$1.5\%$ & $(1.86\pm1.12)\%$ & \checkmark \\
$D^0\to K^+K^-$ & SCS, $T$ & $0.7\%$--$0.8\%$ & $(7.7\pm5.7)\times10^{-4}$ & \checkmark \\
$D^0\to\pi^+\pi^-$ & SCS, $T$ & $0.6\%$--$0.7\%$ & $(23.2\pm6.1)\times10^{-4}$ & \checkmark \\
$D^0\to\pi^0\pi^0$ & SCS, $E+T$ & $0.5\%$--$1.0\%$ & $<5\%$ & \checkmark \\
$D^+\to\pi^+\pi^0$ & SCS, $\Delta I=3/2$ & $<0.1\%$ & $<0.3\%$ & \checkmark \\
$D^+\to\bar K^0\pi^+$ & CF & $<0.05\%$ & $<0.4\%$ & \checkmark \\
$D^0\to K^-\pi^+$ & CF & $<0.05\%$ & $<0.1\%$ & \checkmark \\
$D_s^+\to K^+\bar K^0$ & SCS, $T+A$ & $0.1\%$--$0.3\%$ & $(-0.02\pm0.46)\%$ & \checkmark \\
$D_s^+\to\pi^+K^0_S$ & SCS, $E+A$ & $0.2\%$--$0.4\%$ & not measured & --- \\
\hline\hline
\end{tabular}
\end{table}
%%%%%%%%%%%%%%%%%%%%%%%%%%%%%%
\section{Conclusion}
%%%%%%%%%%%%%%%%%%%%%%%%%%%%%

We have investigated the phenomenological effects of scalar diquarks on CP violation in the decay channels $D^0 \to K_S^0 K_S^0,K^+K^-,\pi^+\pi^-$. 
%Within the Standard Model (SM), the direct CP asymmetry in this mode is suppressed due to the vanishing of the leading amplitude in the U-spin limit and the dominance of exchange topologies.
It is well-known that the expected Standard Model asymmetries in these channels remain below the percent level. We have shown that the introduction of a $S_6$ scalar diquark with a mass of order 1 TeV and perturbative couplings can naturally generate CP asymmetries at the percent level, consistent with current experimental observations for $D^0 \to K_S^0 K_S^0$.

A comparative analysis of $S_6$ and $S_3$ scalar diquark representations demonstrates that the $S_6$ scalar diquark provides the most viable explanation of the data. Its symmetric color structure, characterized by the relation $C_1^{\rm NP} = C_2^{\rm NP}$, avoids the color suppression present in exchange amplitudes with $S_3$ scalar diquark. As a consequence, it allows for constructive interference with the SM contribution. This leads to an enhancement of the $D^0 \to K_S^0 K_S^0$ decay amplitude and of the corresponding CP asymmetry. In contrast, the color-triplet representation, which implies $C_1^{\rm NP} = -C_2^{\rm NP}$, induce  destructive color interference and its effect on  $D^0 \to K_S^0 K_S^0$ CP asymmetry is suppressed. 

Moreover, the flavor-dependent structure of the  diquark couplings, which allows to have the hierarchy $\lambda_{ud} > \lambda_{us}$, provides a coherent description of CP violation across multiple charm decay channels. This hierarchy enhances the asymmetry in $D^0 \to K_S^0 K_S^0$ while  reproducing the observed pattern of positive direct CP asymmetries in $D^0 \to K^+ K^-$ and $D^0 \to \pi^+ \pi^-$. In addition, it  can generate the observed deviation from the U-spin sum rule.

As a conclusion, the $S_6$  scalar diquarks are a  well-motivated candidates for explaining enhanced CP violation in charm decays. Such states arise naturally in  grand unified theories based on gauge groups such as $SO(10)$ and $E_6$. Future precision measurements of CP asymmetries in charm decays, as well as direct searches for scalar diquarks at high-energy colliders, will provide important tests of this scenario and further clarify the origin of CP violation in the charm sector.

%%%%%%%%%%%%%%%%%%%%%%%%%%%%%%%%%%%%%%%%%%%%%%%
\section*{Acknowledgements}
We dedicate this work to the memory of our esteemed colleague, Dr. Gaber Faisel, who sadly passed away at the beginning of this project. We are deeply grateful for his foundational insight regarding the diquark contributions to the $D \to K_S K_S$ decay, which motivated this research. The work of S.~K.~is partially supported by Science, Technology $\&$ Innovation Funding Authority (STDF) under grant number 48173. The work of D.D. is supported by Secretaria de Ciencia, Humanidades, Tecnologia e Innovaci\'on (SECIHTI) and Sistema Nacional de Investigadoras e Investigadores (S.N.I.I.), Mexico.

%\bibliographystyle{apsrev4-2}
%\bibliography{ref.bib}
%\include{bib}



\end{document}